\documentclass[trackchanges, twocolumn]{aastex701}
\usepackage{CJKutf8}

\begin{document}
\title{Three-Dimensional Dust Distribution in the Jovian System from Juno/Waves Observations: Insights into the Halo Ring and Magnetospheric Dust}

\author[orcid=0009-0004-5316-4487,gname=Yuqi,sname=Zhang]{Yuqi Zhang
 \begin{CJK*}{UTF8}{gbsn}
(张雨麒)
\end{CJK*}}

\affiliation{Department of Earth and Space Sciences, Southern University of Science and Technology, Shenzhen, China}
\email{yqz26@bu.edu}  

\correspondingauthor{Shengyi Ye}

\author[orcid=0000-0002-3064-1082,gname=Shengyi,sname=Ye]{Shengyi Ye\begin{CJK*}{UTF8}{gbsn}
(叶生毅)
\end{CJK*}} 

\affiliation{Department of Earth and Space Sciences, Southern University of Science and Technology, Shenzhen, China}
\email[show]{yesy@sustech.edu.cn}

\author[orcid=0009-0001-1938-8450,gname=Yuting,sname=Li]{Yuting Li\begin{CJK*}{UTF8}{gbsn}
(李禹廷)
\end{CJK*}}
\affiliation{Department of Earth and Space Sciences, Southern University of Science and Technology, Shenzhen, China}
\email{12510209@mail.sustech.edu.cn}

\author[orcid=0009-0007-6827-5819,gname=Wenyue,sname=Li]{Wenyue Li\begin{CJK*}{UTF8}{gbsn}
(李文跃)
\end{CJK*}}
\affiliation{Department of Earth and Space Sciences, Southern University of Science and Technology, Shenzhen, China}
\email{12531164@mail.sustech.edu.cn}

\author[orcid=0000-0003-2480-9851,gname=Guangzhou,sname=Wang]{Guangzhou Wang\begin{CJK*}{UTF8}{gbsn}
(王广洲)
\end{CJK*}}
\affiliation{Department of Earth and Space Sciences, Southern University of Science and Technology, Shenzhen, China}
\email{12232668@mail.sustech.edu.cn}

\author[orcid=0009-0008-8942-1720,gname=Xinya,sname=Duanmu]{Xinya Duanmu\begin{CJK*}{UTF8}{gbsn}
(端木新亚)
\end{CJK*}}
\affiliation{Department of Earth and Space Sciences, Southern University of Science and Technology, Shenzhen, China}
\email{12331084@mail.sustech.edu.cn}


\begin{abstract}
Discoveries regarding the dusty rings of Jupiter and the Galilean satellites' dust environment have been continuously refined by orbiters and flybys. Leveraging Juno Waves instrument electric field data, we developed a hybrid recognition framework, coupling \citeauthor{Kvammen2023}’s Convolutional Neural Network (CNN) with rule-based differential peak analysis, to systematically map the Jovian dust environment. This automated pipeline successfully identified over 150,000 dust impacts, effectively isolating dust signals from intense magnetospheric noise, providing a high-resolution catalog of Jovian microdust distribution and offering a robust technical foundation for future missions. Analysis of the vertical cross-section of the Jovian halo ring reveals a more detailed dust distribution structure, with a distinct number density enhancement near the center of the halo ring. Moreover, we report the continued evidence of dust populations near or in the Jovian magnetosheath through identification of background magnetic and plasma data instant variations during magnetospheric boundary crossings.


\end{abstract}


\section{Introduction}
Since the 1970s, in situ exploration of the dust environment within the Jupiter system has progressively advanced our understanding of planetary rings, satellite ejecta, and magnetospheric dust dynamics. Early measurements by Pioneer 10 and Pioneer 11 first detected dust particle signatures near Jupiter, suggesting the existence of a complex circumjovian dust environment \citep{SINGER1976197, Zook1982}. Subsequently, Voyager 1 and Voyager 2 directly imaged Jupiter’s faint ring system in 1979, leading to its discovery \citep{Smith1979}, while plasma wave observations revealed transient electric signals \citep{Tsintikidis1996} generated by hypervelocity dust impacts on the spacecraft, establishing a new technique for dust detection using electric antennas \citep{Aubier1983,Gurnett1983}. The Galileo mission later provided the first long-term in situ investigation of Jovian dust \citep{Grun1996}, demonstrating that volcanic activity on Io is the dominant source of high-speed nanometer-sized dust streams accelerated by electromagnetic forces within Jupiter’s magnetosphere \citep{Graps2000}. Galileo observations also revealed the fine structure of the main ring, identified impact ejecta from small inner satellites as the primary source of ring material \citep{Burns1999, OckertBell1999}, and discovered tenuous dust clouds surrounding the Galilean satellites \citep{Kruger2003}. Cassini–Huygens confirmed Io’s major contribution to the Jovian dust population and detected possible ejecta from Europa and other satellites \citep{Postberg2006}. New Horizons contributed to the understanding of previously unresolved fine structures within Jupiter’s main ring \citep{Showalter2007}, and Ulysses provided complementary measurements of dust transport in the heliosphere \citep{Grun1993}. More recently, Juno has enabled a new era of long term dust investigations. Without direct dust detector, its Waves instrument continuously records transient electric field signals generated by hypervelocity dust impacts \citep{ye2020}, while the magnetometer (MAG) associated Advanced Stellar Compass detects secondary ejecta produced by dust collisions with the spacecraft structure \citep{Benn2017}. These observations provide critical constraints on the spatial distribution, source mechanisms, and magnetospheric transport processes of dust populations throughout the Jovian system.

Juno has been in operation for 15 years and orbiting Jupiter for 10 years. We identified all the dust impact signals from orbit 0 to orbit 71, especially each available perijove (PJ). By a joint method combining machine learning and differential peaks, we selected the waveforms containing dust signals from over two million original snapshots. Then we characterize the dust distribution within the Jovian system, providing insights into halo ring, satellites and their dusty environment.

\section{Methodology}
\subsection{Detection Mechanism}

Electric field antennas can detect dust impacts through transient plasma signals generated during hypervelocity collisions between dust particles and spacecraft surfaces. When a dust grain impacts a solid target at velocities ranging from several to hundreds of kilometers per second, a fraction of its kinetic energy is converted into impact ionization, producing a rapidly expanding plasma cloud composed of electrons, ions, and neutral fragments. Laboratory impact experiments\citep{Zaslavsky2015, Nouzak2018} and Numerical simulations\citep{Shen2021, Shen2023} have demonstrated that the released charge strongly depends on both the particle mass and impact velocity, commonly following empirical power-law scaling relations. It is further shown that the subsequent evolution of the impact plasma cloud, including charge separation, recollection, and expansion dynamics, is controlled by the spacecraft potential, antenna geometry, and surrounding plasma environment. These processes induce short-duration perturbations in the local electric field and spacecraft floating potential, which can be recorded as impulsive voltage signatures by onboard electric field antennas. The morphology of the detected waveform, such as its amplitude, polarity, rise time, and spectral characteristics, contains information about the impacting dust population and ambient plasma conditions. Inversely, statistical analyses of these transient signals therefore enable indirect estimation of dust impact rates, number density, and size distributions. Owing to their high sensitivity to micrometer-scale particles, antenna-based techniques have become an important approach for space dust detection, and their physical interpretation has been validated through extensive laboratory studies, numerical modeling, and multiple in situ observations in planetary and interplanetary environments\citep{Tsintikidis1996, Gurnett1983, Malaspina2016, Szalay2024, ye2020}.

\subsection{Instrument and Data}

The Waves instrument onboard Juno is designed to investigate radio emissions and plasma waves within Jupiter’s magnetosphere \citep{Kurth2017}. The instrument consists primarily of a V-shaped electric dipole antenna with a tip-to-tip length of approximately 4 m, a magnetic search coil, two preamplifiers, three receivers, and a digital processing unit. Waves measures electric field fluctuations over a frequency range from 50 Hz to 45.25 MHz and magnetic field fluctuations from 50 Hz to 20 kHz. Electric field measurements are obtained from the potential difference between two antenna elements in dipole mode, which provides improved suppression of spacecraft-generated noise compared with monopole measurements that detect the potential difference between a single antenna and the spacecraft body\citep{ye2019cassini}.

Hypervelocity dust impacts on the spacecraft surface generate transient plasma clouds that perturb the local electric field and spacecraft potential, producing impulsive voltage signatures detectable by the Waves antennas. Previous studies conducted by \citet{ye2020} have shown that dust signals can still be detected even when the antennas are partially blocked by the spacecraft structure or solar panels, indicating that the observed signals are primarily produced by impact-generated plasma associated with collisions on the spacecraft body. The effective area, defined as the projected area perpendicular to the spacecraft velocity vector, is commonly used to estimate dust fluxes and number densities. Due to variations in spacecraft attitude, the effective area of Juno changes with time and location, typically ranging from 20 to 40 m$^{2}$. In this study, a representative effective area of 30 m$^{2}$ is adopted.

The Waves data products are archived in ASCII table format within the NASA Planetary Data System (PDS), together with standard label files containing acquisition time, instrument operating mode, calibration coefficients, unit information, and scientific measurements. Because dust impact signals generally occur on millisecond timescales, high-time-resolution waveform data are required for reliable identification and analysis. Therefore, this study uses low-frequency burst waveform data acquired by the Low Frequency Receiver (LFR) in burst mode, downloaded from the PDS archive. The selected dataset has a sampling rate of 50 kHz, and each waveform record snapshot spans 122.88 ms with 6144 samples. The analyzed interval covers observations from 4 July 2016 (DOY 184) to 4 April 2025 (Day of Year, DOY 94), comprising a total of 2,145,778 burst waveform snapshot records.

\subsection{Dust Impact Signal Identification}

\begin{figure}[htbp]
    \centering
    \includegraphics[width=1\linewidth]{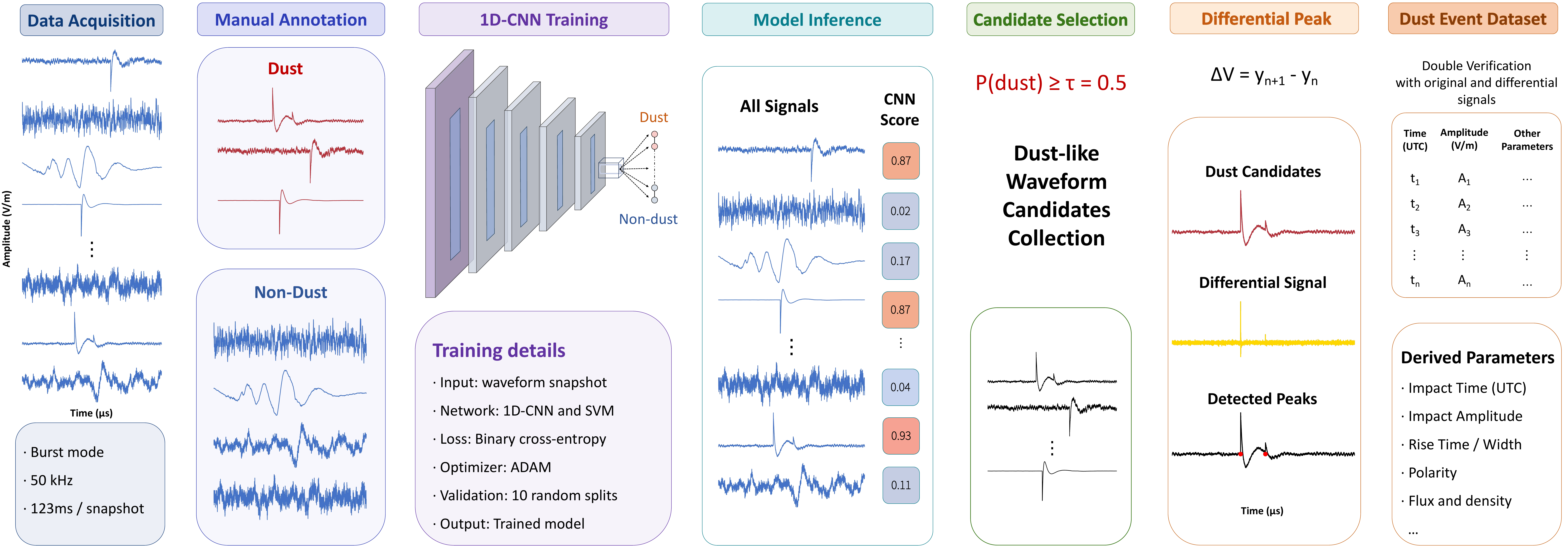}
    \caption{A demonstration of the dust signal selection work flow based on waveform}
    \label{fig_pipline}
\end{figure}

Figure \ref{fig_pipline} illustrates the workflow of our new pipeline, which combines machine learning with traditional methods to establish a dust database.

\begin{figure}[htpb]
    \centering
    \includegraphics[width=0.7\linewidth]{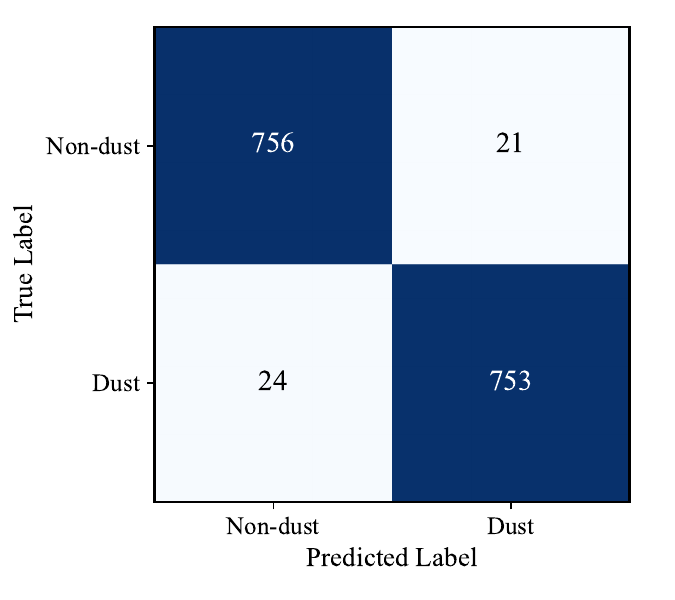}
    \caption{Confusion matrix for the dust event classifier on the independent test dataset. The rows denote the true labels and the columns denote the predicted labels. Most samples are correctly classified (diagonal elements), demonstrating balanced performance for both dust and non-dust classes.}
    \label{fig_confusion_matrix}
\end{figure}

Firstly, to overcome the limitations of traditional threshold-based methods, this study adopts a one-dimensional convolutional neural network (1D-CNN) framework for dust signal identification, originally developed by \citet{Kvammen2023} for the Radio and Plasma Waves (RPW) instrument onboard Solar Orbiter. During preprocessing, waveform snapshots are standardized through offset removal, median-filter denoising, data compression, and amplitude normalization, reducing environmental variations and improving training efficiency. The model is used to classify whether a waveform snapshot contains a dust impact signal, achieving an overall classification accuracy of 97.1\% on an independent test set, with results shown in Table \ref{table_CNN} and Figure \ref{fig_confusion_matrix}. Based on the trained model, 77,040 dust waveform snapshots were identified and subsequently verified through manual inspection.

\begin{deluxetable}{lc}[htbp]
\tablewidth{0pt}
\tablecaption{Performance metrics of the dust signal classifier on the independent test dataset.\label{table_CNN}}
\tablehead{
\colhead{Metric} & \colhead{Value}
}
\startdata
Accuracy                  & \textbf{97.10\%} \\
Precision         & 97.29\% \\
Recall            & 96.91\% \\
F1-score          & 97.10\% \\
\enddata
\end{deluxetable}

Secondly, to accurately determine the timing and amplitude of individual dust impacts, both the original waveform and its first-order difference waveform are analyzed jointly. The first-order difference, defined as $\Delta E[n]=E[n+1]-E[n]$, represents the temporal gradient of the signal and enhances the sharp rising edge characteristic of hypervelocity dust impacts. In contrast, baseline drift, low-frequency fluctuations, and quasi-periodic plasma waves are significantly suppressed in the differential domain, improving edge localization and event separation. Peak detection performed on the differential waveform therefore provides higher sensitivity and robustness for dust event identification. Using this approach, a total of 155,043 dust impact events were identified. Among them, 34,238 clipped signals exhibit saturated signal amplitudes and are excluded from subsequent analysis involving impact amplitude, inferred dust mass, and particle size distributions.

\section{Results}
\subsection{Dust Distribution}

\begin{figure}[htbp]
    \centering
    \includegraphics[width=1\linewidth]{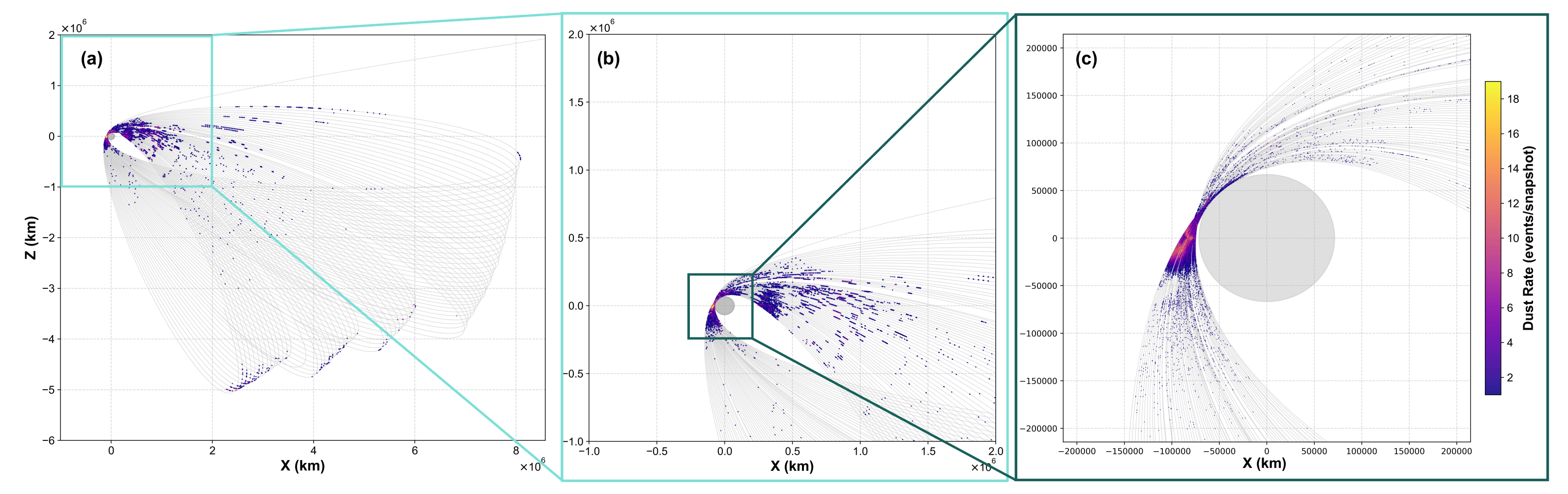}
    \caption{The overall distribution of dust impact events detected by Juno is shown as a colored scatter plot in Jupiter Equatorial Inertial frame (JEIJ2000)\citep{Wang2023Jupiter}. The spacecraft trajectories are shown in gray, and the gray ellipsoid represents the Jupiter itself. Spacecraft and planetary ephemerides were computed using the NAIF SPICE toolkit through the Python wrapper SpiceyPy \citep{Annex2020}, together with the corresponding SPICE kernels provided by NAIF.}
    \label{fig_all_dust}
\end{figure}

Overall, the identified dust population in Figure \ref{fig_all_dust} is broadly consistent with previous direct observations and numerical simulations of the Jovian dust environment. Persistent dust impact signals are observed within the Jovian halo ring region, near the Galilean satellite orbits, and over Jupiter’s northern polar region, indicating the presence of substantial dust populations in these regions. Inside approximately 3 $R_\mathrm{J}$ (1 $R_\mathrm{J}$ = 71488 km), dust impacts are detected over nearly the entire sampled spatial domain. In contrast, no significant dust enhancement is identified in the southern relatively high latitude region. This absence is related to the larger spacecraft distance from Jupiter during southern polar passes, since most polar dust detections in the northern hemisphere are concentrated within about 1 $R_\mathrm{J}$ above the planet's surface.

Due to the highly asymmetric orbit of Juno, symmetric equatorial coverage within 10 $R_\mathrm{J}$ is not available, limiting direct north–south comparisons in the inner system. Nevertheless, during intervals with comparable spatial coverage in northern and southern hemispheres, the dust distribution exhibits weak asymmetries between both regions. These features may reflect intrinsic dynamical processes, although observational biases associated with spacecraft trajectory and viewing geometry cannot yet be fully excluded. Continued observations during the second extended mission of Juno are expected to further improve the spatial completeness of the Jovian dust distribution and help clarify the origin of these asymmetries.

In particular, enhanced dust concentrations are identified near the orbit of Io, consistent with its role as the dominant dust source in the Jovian system. Dust impact signatures are also detected near the orbital regions of Europa and Ganymede, suggesting additional contributions from impact-generated ejecta associated with these satellites.

As the orbit of Juno evolved over successive PJ passes, intermittent dust signals were observed in the low-latitude southern hemisphere beyond approximately 20 $R_\mathrm{J}$. In contrast, dust detections at high latitudes are considerably less frequent, indicating that the outer dust population is primarily concentrated near the Jovian equatorial plane.

Notably, during the early orbits of the Prime Mission, when the apojove trajectory of Juno was located on the dawnside of Jupiter, as well as after the beginning of the Extended Mission when apojove shifted toward the duskside, repeated dust detections were recorded at large radial distances from Jupiter. These observations indicate that dust is widely distributed throughout the outer Jovian system, with pronounced spatial identities.

\subsection{Halo Ring}
The overall micron- and submicron-scale dust distribution within the inner Jovian system is generally consistent with the known morphology of the Jovian ring system, especially the halo ring population previously identified by imaging observations \citep{OckertBell1999,dePater2008} and dynamical simulations \citep{horanyi2010plasma}. The majority of dust impacts are concentrated close to the Jovian equatorial plane, while detections at higher latitudes become increasingly sparse. However, several previously unresolved spatial features are revealed by the long-term Juno/Waves dust survey, which provides a vertical cross-section observation of the Halo ring.

\begin{figure}[htbp]
    \centering
    \includegraphics[width=1\linewidth]{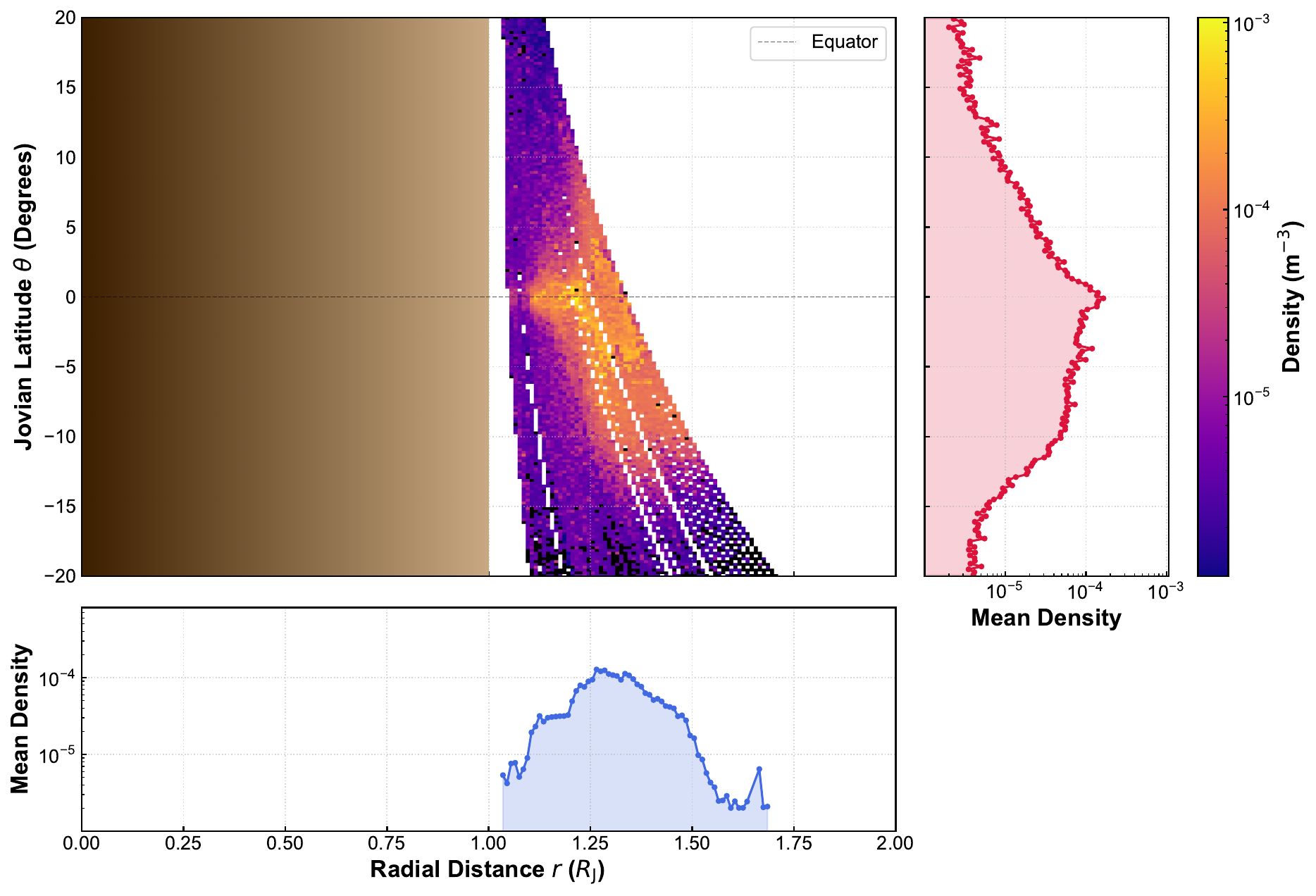}
    \caption{The cumulative dust number density distribution of the halo ring is displayed in r--$\theta$ polar coordinates. Each bin with a size of 0.01 R$_{\text{J}}$ and 0.2 degree represents the average dust number density derived from all available dust impact observations and calculations. Black regions indicate intervals when Waves was operating but no dust impacts were detected, whereas blank regions correspond to periods when Waves was not operating or when Juno did not pass through the region. The lower panel shows the radial profile averaged over latitude, while the right panel shows the latitudinal profile averaged over radial distance. The brown gradient-shaded region represents Jupiter. }
    \label{fig_dust_ring_density}
\end{figure}
In particular, the halo-associated dust population exhibits a substantially broader vertical extent than predicted by radial confined torus models. Based on Juno's flybys through the halo ring region, in situ dust measurements suggest a broader vertical extent of the halo ring dust population, reaching approximately $\pm(37,500-42,500)$ km. Radially, the observed dust distribution indicates that the halo ring boundary may extend closer to Jupiter, potentially reaching radial distances of $\sim80,000$ km or less.

The detected distribution also shows indications of density heterogeneity and grain sizes related distortions. From Figure \ref{fig_dust_ring_density} of halo ring dust distribution, assuming a circular cross sectional shape, the number density gradually increases towards the center, suggesting that the Jovian dust distribution is spatially structured rather than previously thought. The dust number density presented in Figure \ref{fig_dust_ring_density} is corrected for the antenna attenuation effect. Using the intrinsic dust impact amplitude distribution derived from regions with negligible antenna attenuation, we corrected the observed impact rates to recover the intrinsic collision rates and subsequently derived the dust flux and number density (see Appendix \ref{appendix:dust_size}). Without this correction, regions with stronger antenna attenuation would preferentially miss low-amplitude dust impacts, resulting in an underestimation of the impact rate and a biased spatial distribution of dust density, which is similar to the gain changes of the Cassini RPWS \citep{Ye2014Cassini}.

\begin{figure}[htbp]
    \centering
    \includegraphics[width=1\linewidth]{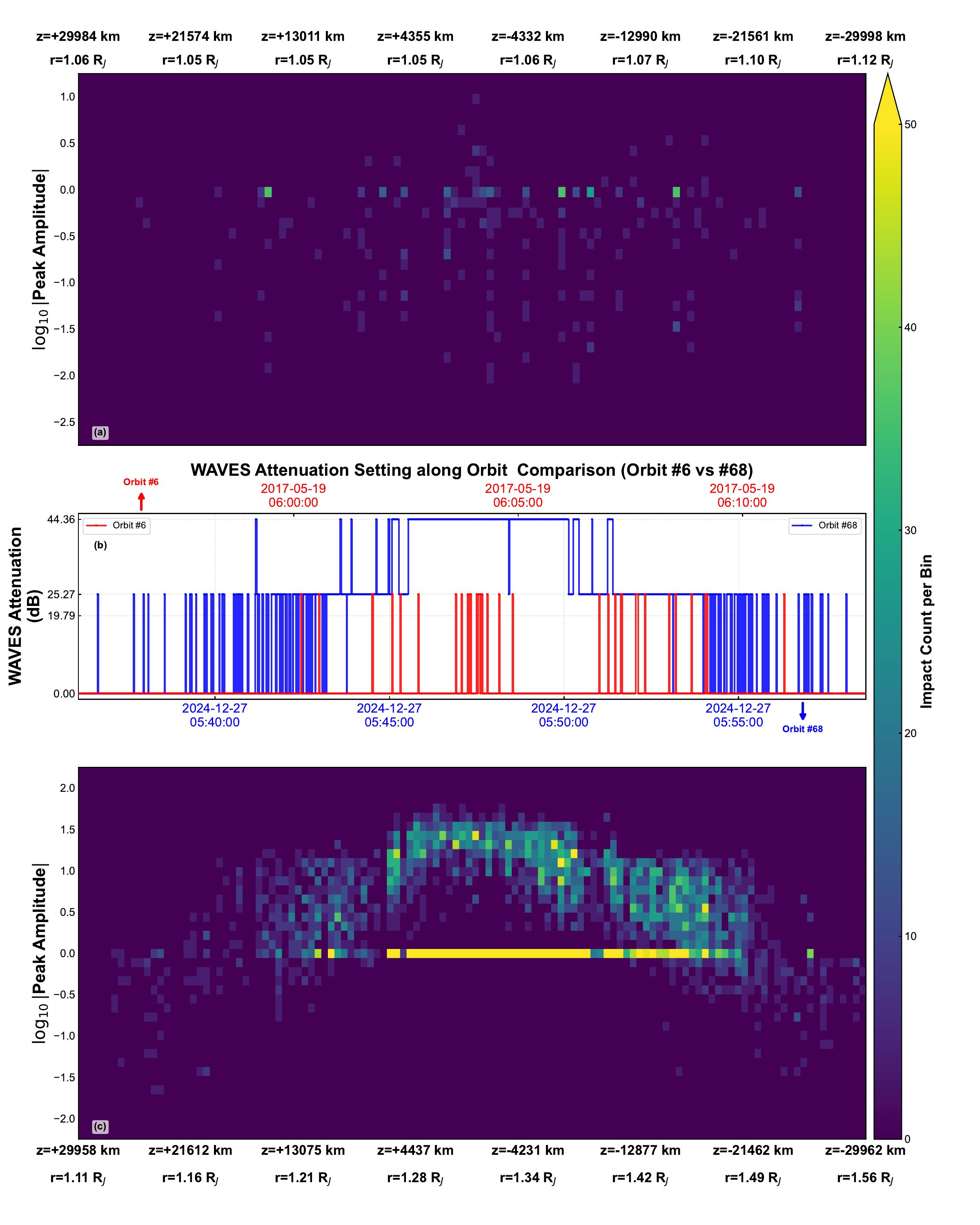}
    \caption{The Comparison of dust impacts detected during PJ 6 and PJ 68. Number of dust impacts recorded in the Waves waveform snapshots as a function of calibrated peak electric field amplitude (base 10 logarithm) and time. Variation of the attenuation of Waves during the PJs.}
    \label{fig_amplitude_spectrogram}
\end{figure}

\begin{figure}[htbp]
    \centering
    \includegraphics[width=1\linewidth]{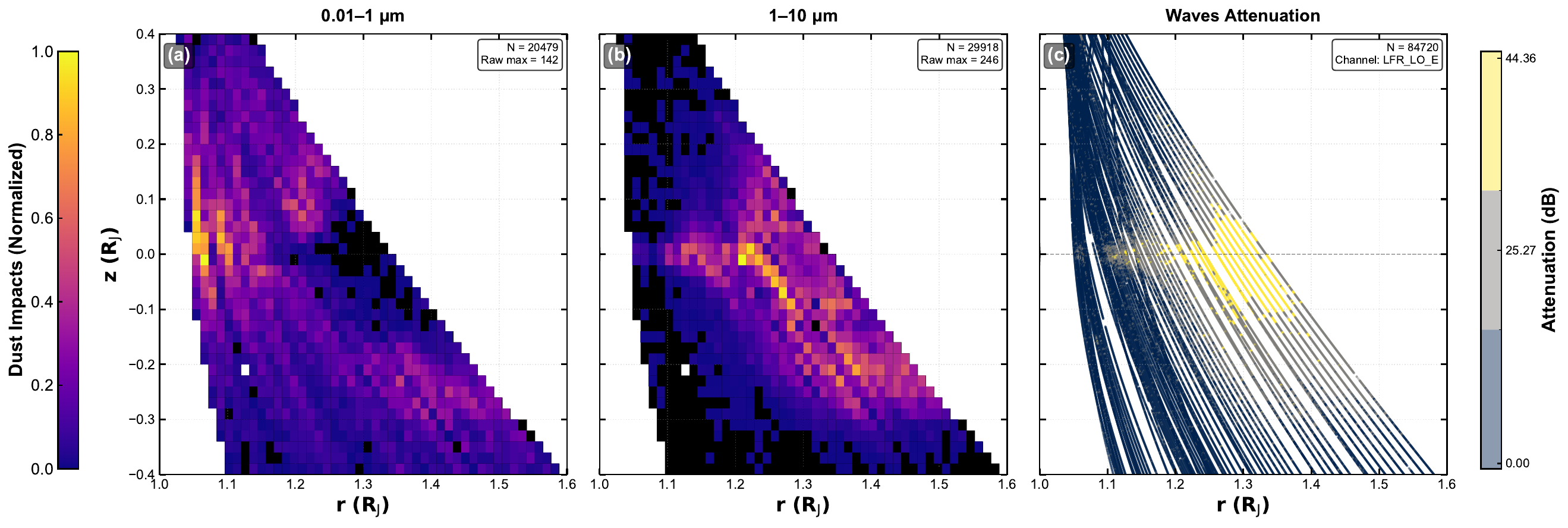}
    \caption{Comparison of the observed dust impacts and Waves attenuation. The spatial bin sizes in (a) and (b) are $\Delta r = 0.012\,R_{\mathrm{J}}$ and $\Delta z = 0.020\,R_{\mathrm{J}}$. The count maps are individually normalized to their respective maxima (color scale: 0--1) to emphasize the relative spatial morphology. (c) shows the attenuation variation in burst mode channel within the halo ring region}
    \label{fig_attentuation_halo}
\end{figure}

Figure~\ref{fig_amplitude_spectrogram} presents the difference of Waves operational state variation between two PJs. During PJ 3, the attenuation level increased only briefly, and the instrument remained close to the zero-attenuation state for most of the observation period. The detected dust impact amplitudes were therefore primarily below 1 V/m. In comparison, PJ 68 experienced higher attenuation states near the equatorial plane, with attenuation increasing by one to two levels. Under these conditions, only higher-amplitude dust impacts were preferentially detected, while weaker signals associated with smaller dust grains were more likely to be missed. This comparison demonstrates that instrumental attenuation introduces a systematic bias in the observed dust amplitude distribution \citep{Ye2014Cassini} and highlights the necessity of applying attenuation correction before deriving dust flux and number density. From Appendix~\ref{appendix:dust_size}, we estimated the corresponding grain sizes based on the signal amplitudes and constructed separate spatial distribution maps for submicron and micron-sized dust populations in Figure~\ref{fig_attentuation_halo}a and \ref{fig_attentuation_halo}b. The number density of larger grains increases from the halo ring boundary toward the center and gradually reaches a plateau, whereas small grains constitute the dominant population near the boundary regions. The comparison between the inferred size distributions and the instrument attenuation states clearly shows that higher attenuation preferentially suppresses the detection of weak impact signals, leading to an underestimation of the dust impact rate. This result highlights the necessity of applying attenuation correction to recover the intrinsic dust population.

\begin{figure}
    \centering
    \includegraphics[width=1\linewidth]{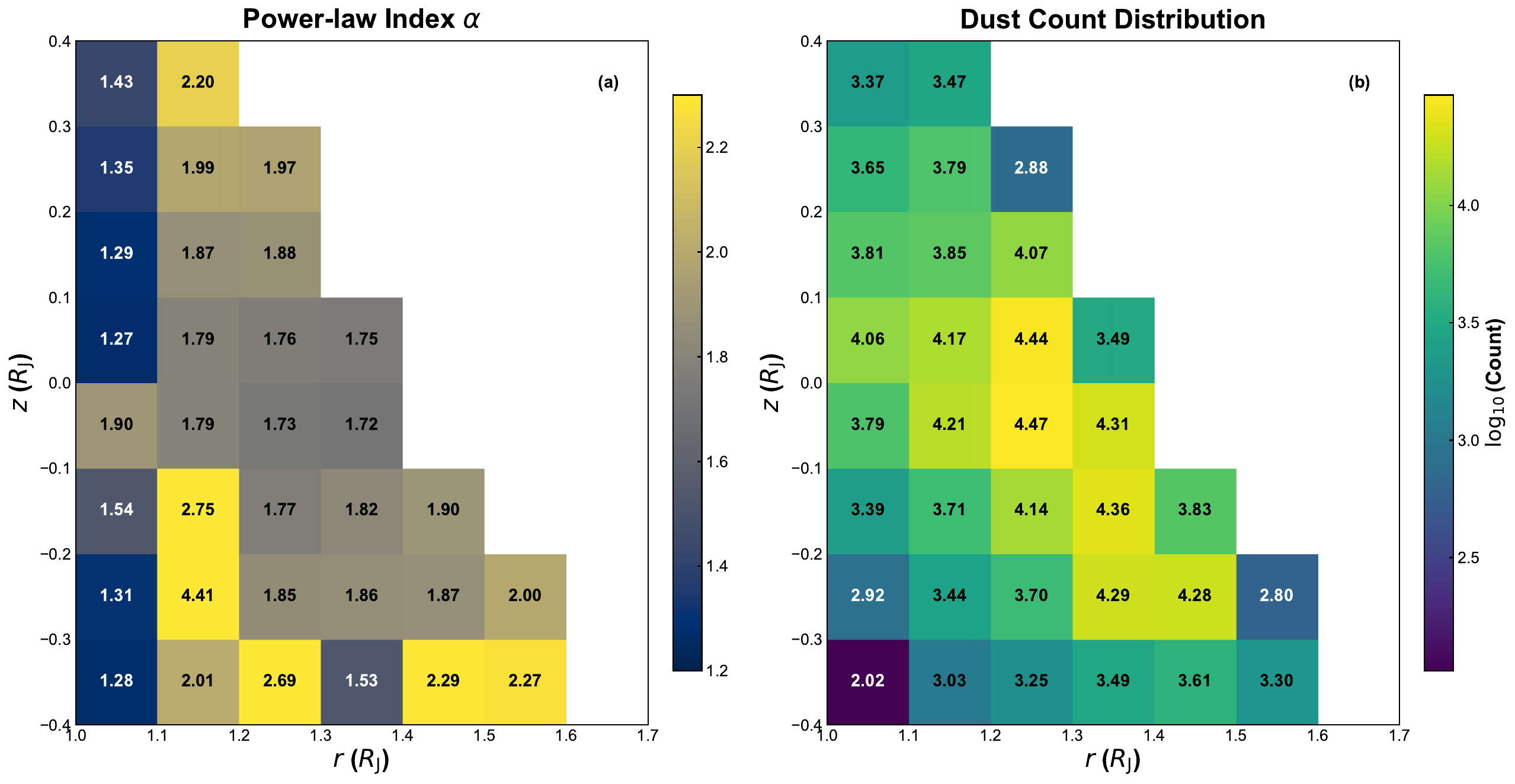}
    \caption{Spatial distribution of the power-law properties of dust impact signals in $0.1R_\mathrm{J} \times 0.1R_\mathrm{J}$ bins. Panel (a) shows the power-law slope $\alpha$ of the dust impact amplitude probability distribution function calculated in each spatial bin. Panel (b) shows the corresponding number of dust impact events per bin displayed on a $\log_{10}$ scale.}
    \label{fig_regional_powerlaw_alpha}
\end{figure}

Previous work by \citet{ye2020} investigated the dust size distribution statistically within an equatorial region of approximately 8000 km. We further extend this analysis by dividing the observed region into finer spatial bins, providing a more detailed view of the spatial variation of dust impact size distributions. Based on the Waves's attenuation and the minimum threshold indicated from Figure~\ref{fig_amplitude_spectrogram}, A statistical analysis of the power law fit, with Maximum Likelihood Estimation (MLE) \citep{Clauset2009powerlaw}, on the dust impact signal peak amplitude probability distribution function is presented in Figure~\ref{fig_regional_powerlaw_alpha}. The number-weighted average of the derived power-law slope is generally consistent with the statistical results reported by \citet{ye2020}. However, a notable decrease in $\alpha$ is observed near the boundary of the halo ring. This variation may indicate a change in the dust impact amplitude distribution in the outer region of the halo ring. Some extremely large values of $\alpha$ are likely associated with bins containing a limited number of dust impacts, where the statistical uncertainty becomes significant and may introduce deviations from the actual distribution.

Taken together, these results indicate that the Jovian dust environment is dynamically complex and likely shaped by the combined effects of satellite ejecta production, electromagnetic transport, plasma interactions, and long-timescale orbital evolution. The observations imply that the circumjovian dust environment may be more complicated than inferred from optical imaging alone, highlighting the importance of long-term in-situ measurements for resolving low-density dust populations. 

\subsection{Dust Environment of the Galilean Satellites}

Jupiter is orbited by a diverse population of satellites, including the four Galilean satellites (Io, Europa, Ganymede, and Callisto), whose orbital distances are 5.9, 9.5, 15, and 26.3 R$\mathrm{_J}$ each. Owing to their large sizes and distinct surface and plasma environments, these satellites play important roles in shaping the Jovian dust environment \citep{lius&chmidt2019}. The Galilean satellites are also expected to contribute to the Jovian dust environment through several processes. Hypervelocity impacts of interplanetary micrometeoroids continuously generate ejecta clouds around their surfaces, while Io additionally produces high-speed dust streams through volcanic activity and subsequent electromagnetic acceleration of charged dust grains \citep{Bagenal2020Io&Europa}. Consistent with this picture, enhanced dust concentrations are identified near the orbit of Io, supporting its long-established role as the dominant dust source in the Jovian system. Proposed contributions from plume activity and sputtering on other three satellites have also been discussed, although their roles remain uncertain. Europa's putative plumes \citep{Roth2014, Sparks2016} were refuted by JWST non-detections and a self-retraction \citep{Roth2026}. Ganymede's bright terrain is debated between cryovolcanic \citep{Solomonidou2026} and tectonic origins, with no plumes detected. Callisto lacks evidence for either, whose surface degradation is exogenic \citep{Schubert2004}. Although localized dust impact enhancements are observed during several close flybys of Europa and Ganymede, the results interpretation is consistent with previous observational and theoretical studies \citep{kurth2023Europa,kurth2022ganymede}, which have not provided conclusive evidence for persistent high-altitude dust plumes reaching hundreds of kilometers or extended ejecta clouds around either Europa or Ganymede. Consequently, the elevated dust fluxes detected during these encounters may instead reflect Juno's interception of corotating dust populations, high-speed dust streams originating from Io, or transient concentrations of interplanetary dust particles, rather than material directly released from the satellite surfaces.

\subsection{Dust Near the Magnetospheric Boundaries}
Beyond inner Jovian dust populations, spatially discrete dust detections are also identified near the apojove portions of the Juno trajectory at radial distances exceeding approximately 60 $R_\mathrm{J}$. These distant detections primarily appear as spatially discrete and relatively low-frequency events concentrated in the low-latitude southern hemisphere.
\begin{figure}[htbp]
    \centering
    \includegraphics[width=1\linewidth]{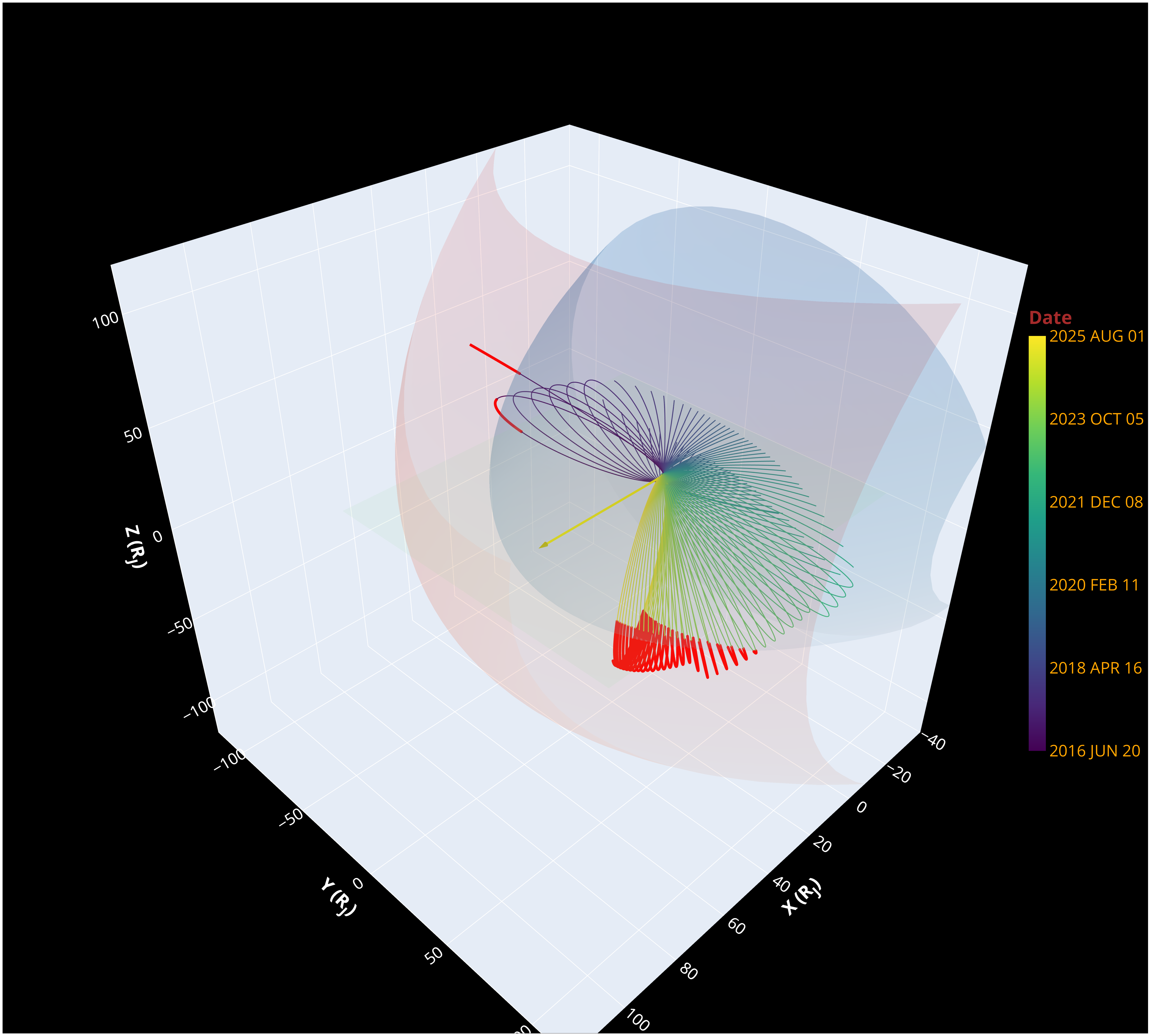}
    \caption{Three-dimensional schematic of the Jovian magnetosphere and the Juno trajectory in the Jupiter-centered and Sun-fixed equatorial frame. Based on geometric of Jovian magnetosphere from \citet{Rutala2025Jupitermag}, the red and blue surfaces with transparency denote the modeled bow shock and the magnetopause, respectively. The Juno trajectory is color-coded according to time. Red curves indicate its flight path outside the magnetopause. }
    \label{fig_dust_magnetosheath}
\end{figure}

\begin{figure}[htbp]
    \centering
    \includegraphics[width=1\linewidth]{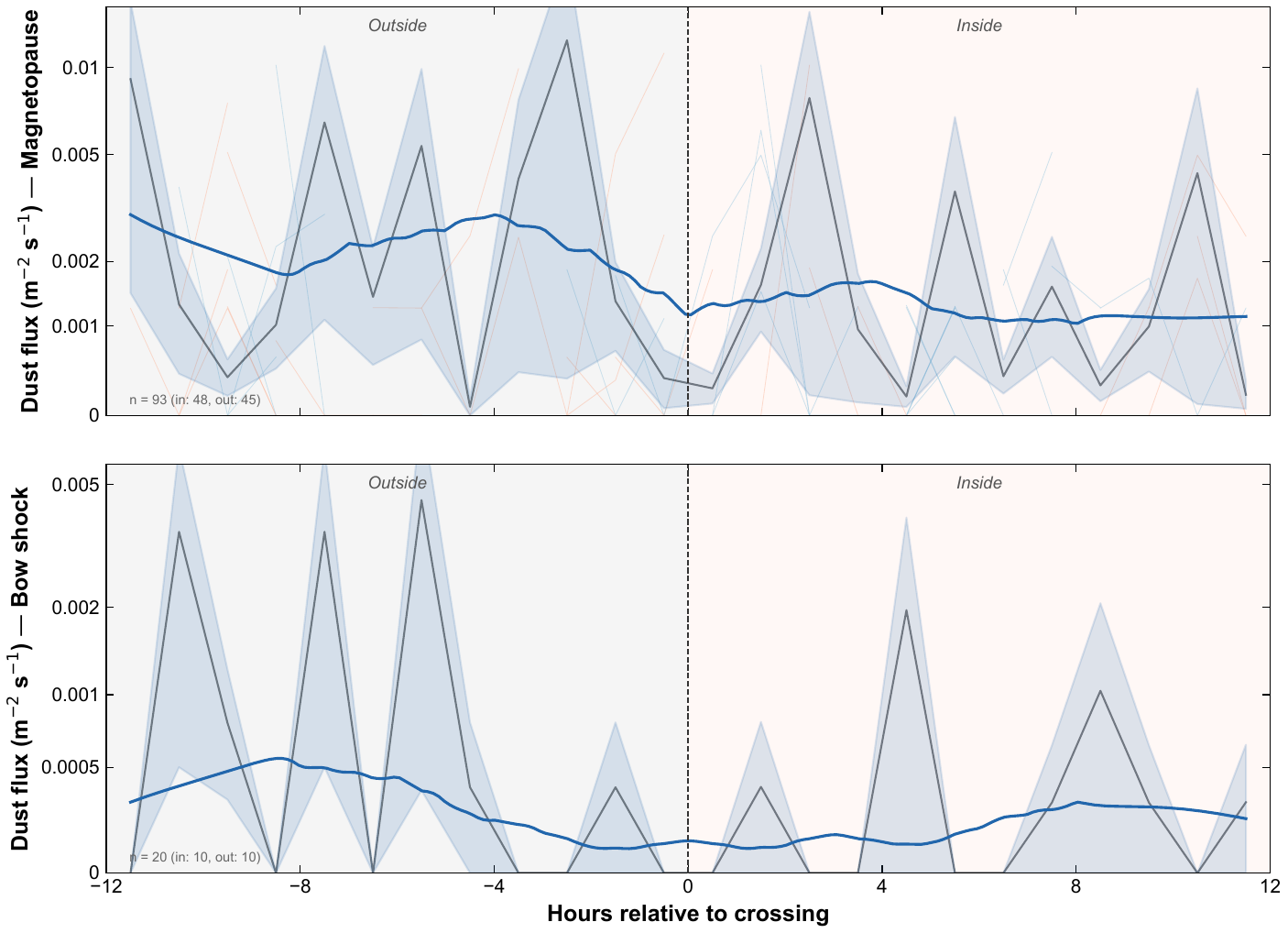}
    \caption{{Dust flux variation across the Jovian magnetopause (top panel) and bow shock (bottom panel). The horizontal axis is centered on the boundary crossing time ($t=0$), with the left and right sides representing the upstream (Outside) and downstream (Inside) regions, respectively. For inbound crossings, the original time sequence is preserved, whereas outbound crossings are reversed to maintain a consistent spatial reference frame. Light blue and light red curves represent the hourly dust flux profiles for individual inbound and outbound crossings, respectively. The black curve shows the arithmetic mean flux of all crossings, and the dark blue curve represents the LOESS-smoothed trend. The shaded region indicates the $\pm1$ standard error of the mean (SEM). The background colors and labels distinguish the Outside and Inside regions. The number of analyzed crossings is indicated in each panel. The dust flux is displayed using an asinh transformation, while the tick labels correspond to the original flux values. }}
    \label{fig_no_flux_change}
\end{figure}

\begin{figure}[htbp]
    \centering
    \includegraphics[width=1\linewidth]{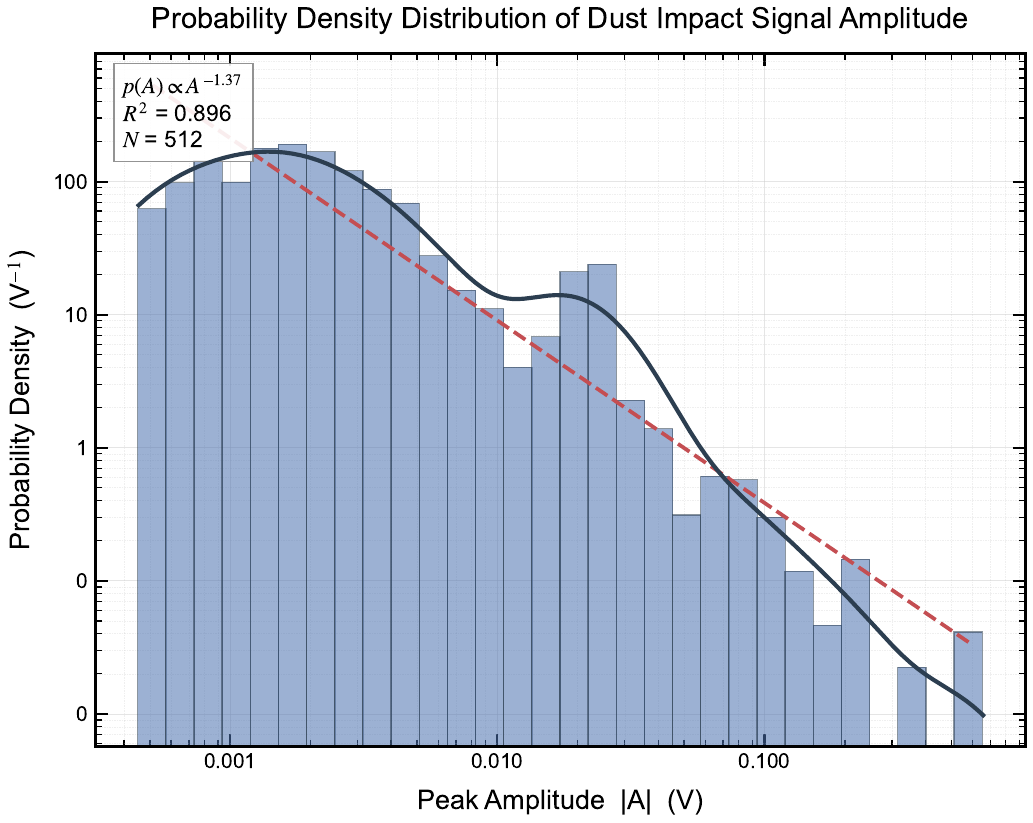}
    \caption{Power-law distribution of peak amplitudes for dust impact events detected near the Jovian magnetospheric boundaries. The log--log probability density distribution is constructed from 512 dust impact events with peak amplitudes ranging from $5\times10^{-4}$ to $0.66~\mathrm{V}$, after excluding events with non-zero clipped fractions and diff-only detections. The solid black curve shows the kernel density estimate (KDE) of the amplitude distribution, while the dashed red line represents the best-fitting power-law model with an exponent of $\alpha=-1.37$.}
    \label{fig_powerlaw_magnetosheath}
\end{figure}

Importantly, similar dust signatures are repeatedly observed during multiple orbital passes, suggesting that they are unlikely to result from isolated instrumental artifacts or transient contamination events. The detections occur during both the early Prime Mission, when Juno apojove was located on the dawnside of Jupiter, and during later Extended Mission phases after apojove migrated toward the duskside (Figure \ref{fig_powerlaw_magnetosheath}). To determine Juno's location within the Jovian magnetosphere, we manually identified the spacecraft's magnetopause and bow shock boundary crossings using in situ magnetic field measurements from MAG and electric field spectrograms (LFR, ~50 Hz–10 kHz, 36 channels) from the Waves instrument, following the methodology of \citet{louis2023effect}. A total of 782 magnetopause and bow shock crossings were identified, including 464 extended crossings results beyond \citeauthor{louis2023effect}. Dust impacts were detected during 112 crossing intervals, comprising 92 magnetopause crossings and 20 bow shock crossings. Among the limited number of magnetopause and bow shock crossings analyzed in this case study in Figure \ref{fig_no_flux_change}, no statistically significant variation in the dust impact flux is observed across either boundary. The apparent periodic modulation in the dust flux does not represent intrinsic temporal variability of the dust population, but primarily results from the operational schedule of the Waves instrument. This result is consistent with previous studies of \citet{grun1996constraints}, suggesting that the detected dust particles are predominantly larger than 0.1 $\mathrm{\mu}$m. 

Based on the identified apojoves intervals, dust impact signals recorded near or within the magnetosheath were further extracted for analysis. Their amplitude distribution is dominated by 10$^{-3}$ -- 10$^{-2}$ V level signals with a power law $\alpha$ = 1.37 as shown in Figure \ref{fig_powerlaw_magnetosheath}, corresponding to impact velocities of approximately 10$^{2}$ km/s. The events are primarily concentrated at relatively low latitudes, with a smaller number of detections at higher latitudes. The low-latitude events are consistent with a possible contribution from high velocity dust associated with Io, whereas the higher-latitude events may include interplanetary dust and other components. 

The observations therefore suggest the possible existence of an extended and dynamically structured outer Jovian dust population that is not fully captured by current steady-state ring models. Possible origins include long-lived satellite ejecta, electromagnetically transported dust grains, or particles influenced by solar radiation pressure and large-scale magnetospheric transport processes.

\section{Discussion}
The present study provides a new in situ characterization of the Jovian dust environment using nearly nine years of Juno/Waves burst-mode observations. Overall, the identified dust distribution is broadly consistent with previous direct measurements and dynamical simulations of the Jovian ring–satellite system, confirming the existence of persistent dust populations associated with the halo ring, the Galilean satellites, and the polar regions of Jupiter. Dust impact signatures were detected throughout most regions inside $\sim3~ R_\mathrm{J}$, indicating that the inner Jovian system is permeated by abundant micron- and submicron-scale particles.

One notable feature revealed by the observations is the broad spatial extent of the halo ring associated dust population. Dust associated signals were detected beyond the conventionally defined outer boundary of the halo ring, implying that the effective radial extent of the Jovian ring environment may be larger than inferred from optical imaging alone. The central region of the halo ring exhibits a great contribution from larger dust grains, suggesting an overall increase in the dust population density in this region. In contrast, the outer boundary regions are dominated by smaller dust grains. With similar mass density, this spatial variation in grain size distribution may result from the charge-to-mass ratio dependent response of dust particles to different forces, including gravity, the Lorentz force, and the corotation electric field \citep{horanyi2010plasma}. In addition, Dust number density local enhancement is identified near the equatorial plane inside the nominal inner boundary of the halo ring. Although its origin remains uncertain, this feature may indicate either an additional dust ring-like component or an inward extension of the halo ring population. Since the Waves instrument provides an in situ measurement of particle impacts rather than scattered light intensity, these results likely reflect the presence of dynamically evolved low optical depth dust populations that are difficult to detect remotely. 

Enhanced dust occurrence rates were also identified near the orbit of Io, supporting the long-established interpretation that Io is a dominant dust source within the Jovian system \citep{lius&chmidt2019}. Dust signals were additionally observed near Europa and Ganymede, indicating that impact-generated ejecta clouds or surface sputtering processes may contribute to local dust populations around these satellites. The spatial morphology of these dust enhancements appears to vary with Juno’s orbital geometry, suggesting that viewing geometry and magnetospheric transport processes both influence the observed distributions.

Another intriguing result is the intermittent detection of sparse dust populations at large radial distances ($>60~R_\mathrm{J}$), particularly in low-latitude regions of the southern hemisphere near apojove. These distant dust detections became more apparent during the early Prime Mission, when Juno’s apojove was located on the dawn side, and again during the Extended Mission after the apojove gradually shifted toward the dusk sector. Although the origin of these distant particles remains uncertain, the observations may indicate the existence of previously underappreciated outer magnetospheric dust populations \citep{Krivov2002}, transient transport pathways, or long-lived grains influenced by electromagnetic forces. Dust impacts were detected within the Jovian magnetosheath in both dawn and dusk sides. Despite multiple magnetopause and bow shock crossings, no significant change in the dust impact flux was observed across either boundary. The absence of a boundary-related flux enhancement is consistent with previous observational studies, further suggesting that the detected dust grains are sufficiently large to remain insensitive to the complicated plasma environment across these boundaries. The origin of this dust is uncertain. Possible contributors include evolved Io-derived dust, dust released from the irregular satellites \citep{chen2024irregular}, and interplanetary dust particles.

The observations also reveal an apparent north–south asymmetry in portions of the dust distribution. However, because Juno’s highly eccentric polar orbit does not provide symmetric equatorial coverage within $\sim10~R_\mathrm{J}$, part of the observed asymmetry may arise from observational bias. Future observations will provide improved spatial coverage and may help distinguish intrinsic asymmetries from orbital sampling effects.

Several limitations should also be considered. First, the conversion between impact-generated electric signals and physical dust properties remains dependent on assumptions regarding impact velocity, charge production efficiency, and grain composition. Second, signal saturation in a fraction of the identified events limits quantitative analysis of the largest-amplitude impacts. Finally, the current study primarily focuses on statistical spatial distributions, while detailed investigations of grain dynamics, temporal variability, and source attribution will require combined modeling efforts involving plasma transport, ring dynamics, and satellite ejecta processes.

\section{Conclusion}

In this study, we analyzed Juno/Waves burst-mode waveform data obtained between 2016 and 2025 and developed a machine learning assisted dust identification framework combining a 1D-CNN with differential peak detection techniques. A total of 155,043 dust impact events were identified from more than two million waveform snapshots, providing one of the most extensive in situ dust datasets currently available for the Jovian system.

The resulting spatial distribution confirms the presence of persistent dust populations associated with the Jovian halo ring, the Galilean satellites, and Jupiter’s polar regions. Align with previous studies, dust at higher density is identified near the orbit of Io, while additional dust populations were observed near Europa and Ganymede. The observations further suggest that the halo ring-associated dust population exhibits a broader vertical extent and may form a three-dimensional toroidal structure, with a central peak in dust number density within the cross-section. In addition to the inner-system dust structures, sparse but repeatable dust detections were observed at the orbits of  Galilean satellites and large radial distances beyond $20~R_\mathrm{J}$, particularly near apojove regions. These detections may indicate the presence of outer magnetospheric dust populations or long-range dust transport processes that have not been fully characterized previously. Dust impacts are also observed during magnetospheric boundary crossings. 

Overall, the present work demonstrates that electric-field waveform measurements from plasma-wave instruments can serve as an effective tool for long-term in situ dust investigations. The combination of machine learning and waveform-based impact analysis provides a robust approach for extracting dust information from large observational datasets and offers new opportunities for studying dust–plasma interactions, ring dynamics, and satellite-generated ejecta throughout the Jovian system.

\section*{Acknowledgments}
This work was supported by the National Natural Science Foundation of China (Grant No. 42274221), the National Key R\&D Program of China (Grant No. 2022YFF0503800), and the Shenzhen Basic Research Program (Natural Science Foundation) (Project No. JCYJ20250604144307010).

The authors acknowledge the Juno Waves team for providing the data used in this study. 

\section*{Data Availability}

The data used in this study are publicly available from the NASA Planetary Data System (PDS). Juno/Waves burst-mode waveform data can be accessed through the PDS Planetary Plasma Interactions Node

The specific dataset used in this work is available at Juno Waves Burst Waveform Data Archive: \url{https://pds-ppi.igpp.ucla.edu/data/JNO-E_J_SS-WAV-3-CDR-BSTFULL-V2.0/DATA/WAVES_BURST/}.

The SPICE kernels used for spacecraft trajectory and coordinate transformations were obtained from the NASA Navigation and Ancillary Information Facility archive NAIF Juno SPICE Kernel Archive: \url{https://naif.jpl.nasa.gov/pub/naif/JUNO/kernels/}.

The data processing and analysis codes developed for this study are available from the corresponding author upon reasonable request.

\appendix
\section{Essential Formulae used for Derivation and Calculation}
\subsection{Dust Impact Charge and Size Distribution Estimation}
\label{appendix:dust_size}

The released charge associated with a dust impact can be estimated from the measured voltage pulse using the effective capacitance of the antenna system. When a hypervelocity dust particle impacts the spacecraft or antenna surface, its kinetic energy is rapidly converted into thermal energy, producing a localized plasma cloud through impact vaporization and ionization. Based on a simplified dust impact plasma model and a capacitive coupling assumption, the collected charge perturbation $\Delta Q$ is related to the measured voltage variation $\Delta V$ as

\begin{equation}
\Delta Q = C_{\mathrm{eff}}\Delta V ,
\end{equation}

where $C_{\mathrm{eff}}$ represents the effective capacitance of the spacecraft-plasma system.

Following the empirical scaling law derived from laboratory measurements, the impact-generated charge can be related to the dust mass and impact velocity as

\begin{equation}
Q \approx c \cdot m \cdot v^{\beta},
\end{equation}

where $m$ is the dust mass, $v$ is the impact velocity relative to the spacecraft, $c$ is an empirical coefficient, and $\beta$ is typically taken as 3.42. Following \citet{Collette2014}, $c$ is adopted as 0.023.

Assuming a characteristic impact velocity, the dust mass can therefore be estimated from the measured voltage amplitude. The cumulative mass distribution can then be derived from the impact charge distribution. Since the impact charge is proportional to the voltage amplitude ($Q\propto \Delta V$), the amplitude distribution can be used as a proxy for the dust mass distribution.

To estimate the characteristic dust size, the particles are approximated as spherical grains:

\begin{equation}
m=\frac{4}{3}\pi\rho\left(\frac{d}{2}\right)^3 ,
\end{equation}

where $d$ is the dust diameter and $\rho$ is the grain density, assumed to be $2.5-3.0~\mathrm{g~cm^{-3}}$ \citep{grun1985,horanyi2010plasma}. Combining the above relations gives

\begin{equation}
d \propto m^{1/3}\propto (\Delta V)^{1/3}.
\end{equation}

Therefore, the power-law index of the amplitude distribution is related to the size distribution index through

\begin{equation}
\alpha_{\mathrm{Amplitude}} \approx \frac{1}{3}\alpha_{\mathrm{size}} .
\end{equation}

\subsection{Impact Rate, Flux, Number Density and their corrections}
\label{appendix:flux and density}

The dust number density distribution presented in this work is derived from the observed dust impact rate after correcting for the attenuation effect of the Juno/Waves instrument. This appendix describes the conversion from the detected dust impact rate to the dust flux and number density, as well as the correction procedure used to recover the intrinsic impact rate.

The observed dust impact rate is calculated from the number of identified dust impacts within a given time interval:

\begin{equation}
R_{\rm obs}(t)=\frac{N_{\rm impact}(t)}{\Delta t},
\end{equation}

where $N_{\rm impact}$ is the number of detected dust impacts and $\Delta t$ is the effective observation time.

The dust impact flux is obtained by normalizing the impact rate to the effective detection area of the spacecraft:

\begin{equation}
\label{cal_flux}
F_{\rm dust} (t)=\frac{R_{\rm true}(t)}{A_{\rm eff}(t)},
\end{equation}

where $A_{\rm eff}$ is the effective detection area and $R_{\rm true}$ is the attenuation-corrected dust impact rate.

Assuming an relative dust impact velocity $v_{\rm dust}$, the dust number density is derived as:

\begin{equation}
\label{cal_num_density}
n_{\rm dust}(t)=\frac{F_{\rm dust}(t)}{v_{\rm dust}(t)},
\end{equation}

where $n_{\rm dust}$ represents the local dust number density. Therefore, an accurate estimation of the intrinsic impact rate is essential for obtaining the spatial dust density distribution.

The corrected impact rate is expressed as:

\begin{equation}
R_{\rm true}(t)=\frac{R_{\rm obs}(t)}{\eta(Att)},
\end{equation}

where $\eta(Att)$ is the detection completeness under a given attenuation state.

The correction assumes that the intrinsic dust impact amplitude distribution follows a power-law function \citep{horanyi2010plasma,Ye2014Cassini}. The complete distribution is obtained from observations under zero attenuation ($Att=0$ dB):

\begin{equation}
p(A)=kA^{-\alpha},
\end{equation}

where $A$ is the calibrated dust impact pulse amplitude, $k$ is the normalization constant, and $\alpha=1.7$ is adopted from the power-law slope of the non-attenuated dust impact amplitude distribution reported by \citet{ye2020}.

The physical amplitude corresponding to the measured signal is corrected using the Waves voltage correction factor:

\begin{equation}
A_{\rm physical}=C_{\rm Vrms}A_{\rm measured},
\end{equation}

where $C_{\rm Vrms}$ is determined by the receiver attenuation state.

For a fixed detection threshold in ADC space, $A_{\rm ADC,min}$, the corresponding physical detection threshold becomes:

\begin{equation}
A_{\rm min}(Att)=A_{\rm ADC,min}C_{\rm Vrms}(Att).
\end{equation}

As the receiver attenuation increases, the corresponding physical detection threshold increases, resulting in reduced detection efficiency for low-amplitude dust impact signals.

The total dust population without attenuation loss is:

\begin{equation}
N_{\rm true}
=
\int_{A_0}^{A_{\rm max}} kA^{-\alpha}dA ,
\end{equation}

Here $A_0$ represents the minimum amplitude detectable under nominal receiver state and defines the reference population used for correction. Under an attenuation state $Att$, the detected population becomes:

\begin{equation}
N_{\rm obs}(Att)
=
\int_{A_{\rm min}(Att)}^{A_{\rm max}} kA^{-\alpha}dA .
\end{equation}

The detection completeness is therefore:

\begin{equation}
\eta(Att)
=
\frac{N_{\rm obs}(Att)}
{N_{\rm true}}
=
\frac{A_{\rm min}(Att)^{-0.7}-A_{\rm max}^{-0.7}}
{A_0^{-0.7}-A_{\rm max}^{-0.7}}.
\end{equation}

Because the maximum dust amplitude is much larger than the minimum detection threshold, the expression can be simplified as:

\begin{equation}
\eta(Att)
\simeq
\left(
\frac{A_{\rm min}(Att)}
{A_0}
\right)^{-0.7}.
\end{equation}

Using

\begin{equation}
A_{\rm min}(Att)=A_0C_{\rm Vrms},
\end{equation}

the completeness correction becomes:

\begin{equation}
\eta(Att)=C_{\rm Vrms}^{-0.7}.
\end{equation}

Therefore, with $Att$ varying under actual observational conditions, the corrected impact rate is:

\begin{equation}
R_{\rm true}(t)
=
R_{\rm obs}(t)C_{\rm Vrms}^{0.7}(t).
\end{equation}

\begin{figure}[htbp]
    \centering
    \includegraphics[width=0.6\linewidth]{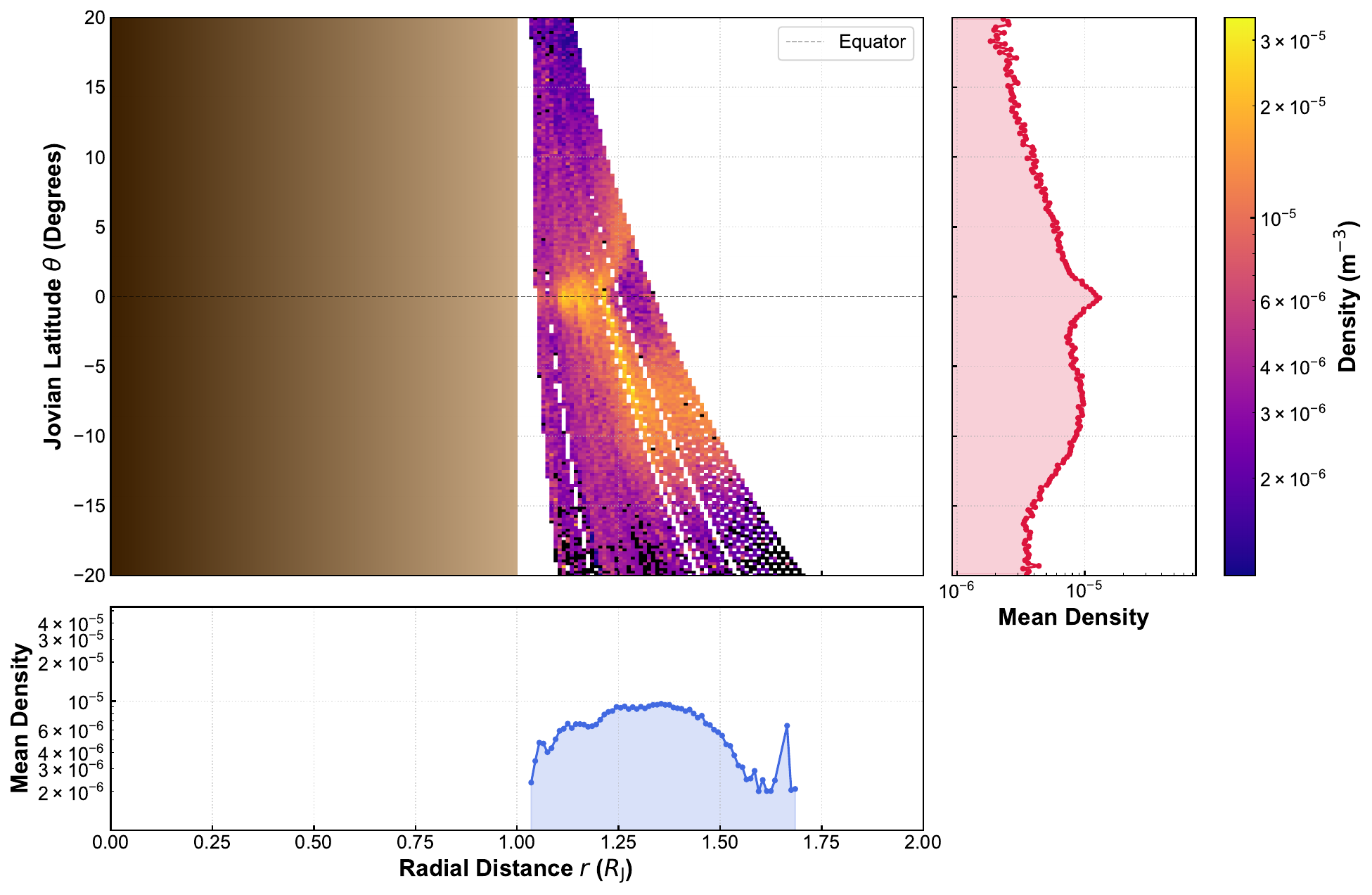}
    \caption{The cumulative dust number density distribution of the halo ring is displayed in r--$\theta$ polar coordinates without correction}
    \label{fig_dust_ring_density_raw}
\end{figure}

The correction factors are applied individually according to the Waves attenuation state. For example, attenuation states of 19.79, 25.27, and 44.36 dB correspond to correction factors of approximately 4.90, 7.52, and 28.6, respectively. However, only three states (0, 25.27, 44.36 dB) appeared during instrument operation as shown in Figure~\ref{fig_attentuation_halo}c. The "cavity"-like structure in Figure~\ref{fig_dust_ring_density_raw} correspond to the attenuation variation. 

After applying the attenuation correction, the corrected impact rate is converted into dust flux and number density following Equations \ref{cal_flux} and \ref{cal_num_density}. This procedure compensates for the loss of weak dust impact signals in highly attenuated observations and prevents systematic underestimation of dust density in regions with reduced instrumental sensitivity.

\section{Examples of Waveform Identification}
\begin{figure}[h]
    \centering
    \includegraphics[width=0.6\linewidth]{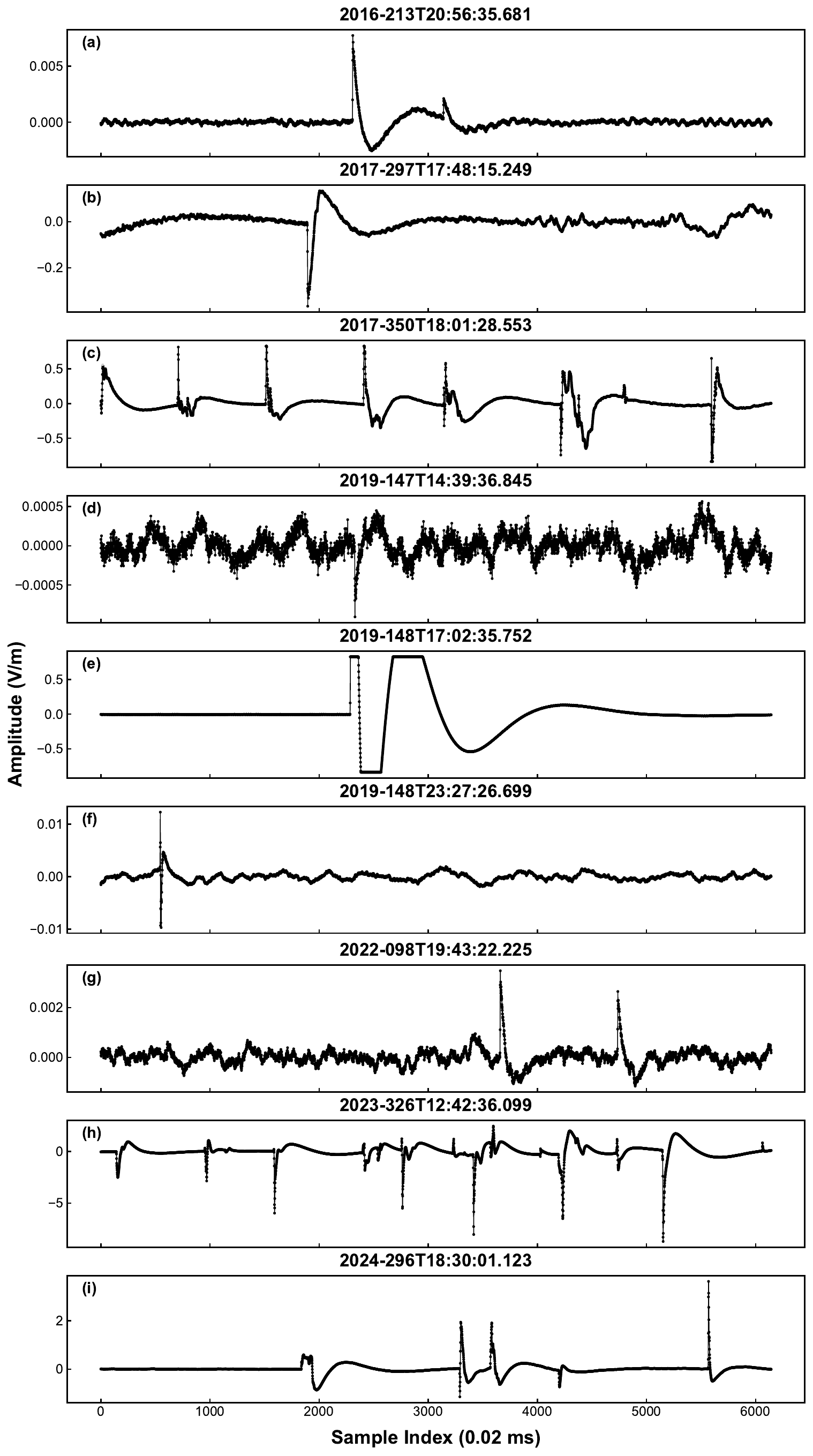}
    \caption{Examples of dust impact waveform snapshots}
    \label{fig_dust_example}
\end{figure}

\begin{figure}[h]
    \centering
    \includegraphics[width=0.6\linewidth]{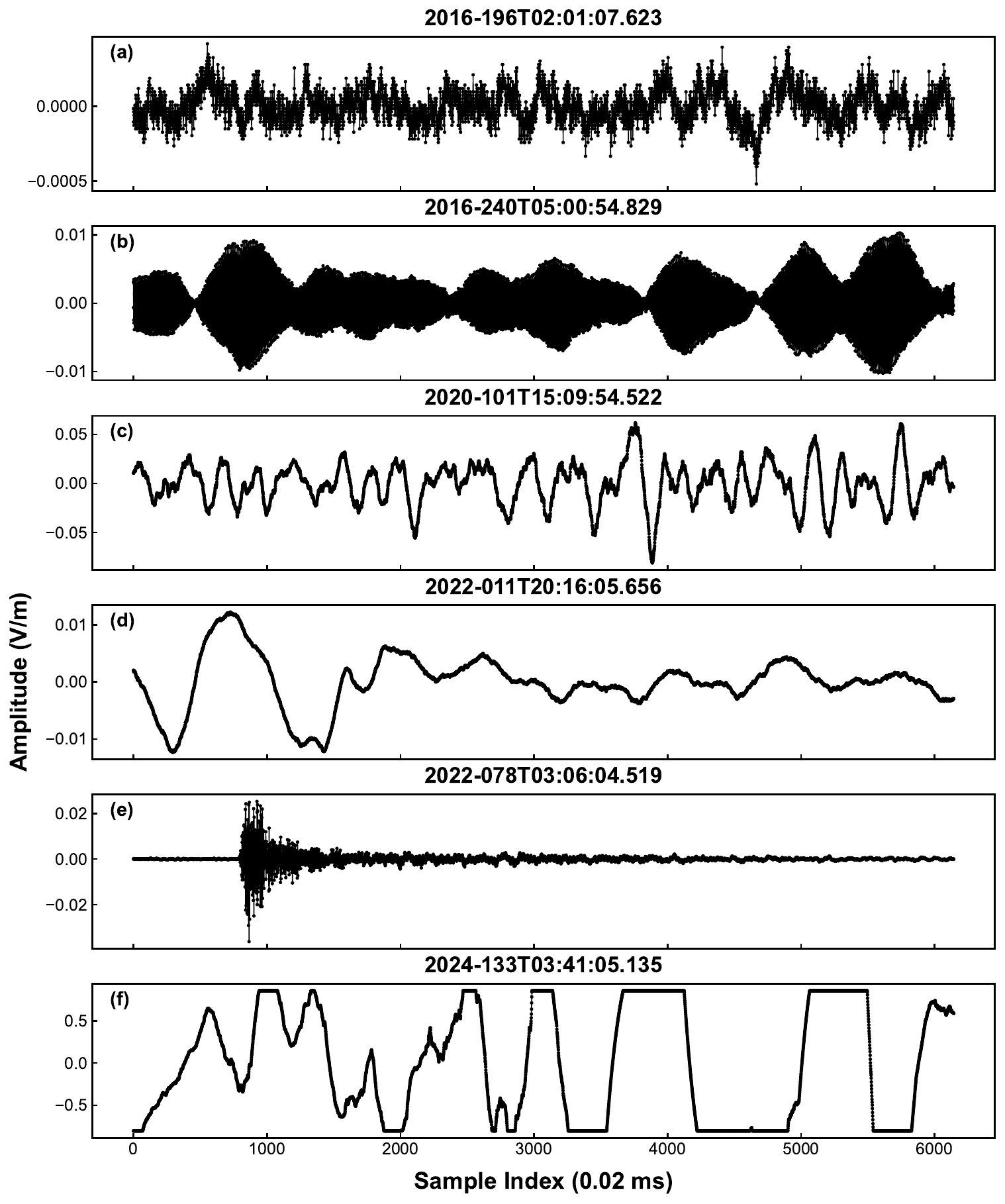}
    \caption{Example of background noise or non dust signal snapshot}
    \label{fig_non_dust_example}
\end{figure}

We manually labeled the waveforms based on their apparent identity. Figure~\ref{fig_dust_example} represents dust impact signal waveform with sharp risings and relaxation damping. More than one dust impact can be recorded in a snapshot. Non dust or background noise waveforms are shown in Figure~\ref{fig_non_dust_example}. 

\clearpage
\bibliography{references}{}
\bibliographystyle{aasjournalv7}
\end{document}